\documentclass[reprint,amsmath,amssymb,aps,twocolumn]{revtex4-1}
\usepackage{graphicx}
\usepackage{dcolumn}
\usepackage{bm}
\usepackage{braket}
\usepackage{hyperref}
\usepackage{subcaption}
\usepackage{caption}
\usepackage{tikz}
\usepackage[framemethod=TikZ]{mdframed}
\usepackage{pgfplots}
\usetikzlibrary{external}
\usepackage{pgfplots}
\pgfplotsset{compat=1.10} 
\usepgfplotslibrary{fillbetween}

\usepackage{tabu}
\begin{document}
\title{Tight-binding model:  correction of the $d$-band approximation}
\author{Jacques R. Eone II } 
\affiliation{%
 \\ \\ 
}%

\date{2021\\} 

\begin{abstract}
The electronic structure, when restricted to the $d$-band approximation, is a computational model that is both efficient and useful for describing transition metals. In the absence of considering delocalized $sp$-states, this approximation gives rise to incorrect surface energies, binding energies, and an inaccurate description of ferromagnetic transition metals.  The present work compares the complexity of implementing corrections with the possibility of using an accurate $sp-d$ approach. Basic force fields based on the second moment approximation continue to be utilized for the description of interactions in transition metals. In contrast, the present study proposes an elementary and more accurate interatomic potential based on hopping parameters depending on distances. The charge distribution and the Stoner model are also analyzed to provide appropriate corrections to the tight-binding picture used to describe ferromagnetic metals and alloys.  
\end{abstract} 

\maketitle

\maketitle

\section{\label{sec:level1}Introduction}
The tight-binding approximation, constrained to the $d$-band and excluding the effect of the $sp-d$ hybridization, remains a popular method for describing transition metals and alloys \cite{cite1,cite2}. This approximation is based on the assumption that the delocalized $sp$-band has a low electron density and does not contribute significantly to the properties of transition metals. The $d$-band described by the approximation is then localized, and the binding and surface energies are not always in agreement with density functional theory (DFT) results. This $d$-band approximation possesses the advantage of reducing the computational cost compared to a tight-binding model, which incorporates all the orbitals. The effect of the $sp-d$ hybridization is of significant importance in the accurate determination of the physical and magnetic properties of transition metals \cite{cite3}. The correlation between the tight-binding approximation, including  the $sp$-band, and the $d$-band approximation is essential for the derivation of appropriate corrections. This correlation has been previously analyzed in another study \cite{cite4}. In that particular study, the magnetic ground state is restored by adding an additional onsite Coulomb repulsion $U_d^{sp}$. This supplementary energy is derived from the interaction between the $sp$-band and the $d$-band. The result of this interaction is a $d$-band that is partially delocalized compared to the initial approximation that does not include $sp-d$ hybridization. The implementation of this correction, which pertains to ferromagnetic metals, should be incorporated into interatomic potentials. In this particular instance, the correction $U_d^{sp}$  attributable to the $sp$-band, is expected to be dependent on the distance. It is evident that an interactomic potential without the $sp$-band is incapable of encompassing the correct electronic structure and deriving accurate physical properties. The second moment approximation (SMA), limited to a single band, is not an adequate methodology for the study of transitional metals.  A derivation of an interatomic potential is possible through utilization of the $sp-d$ basis set, such as the NRL-tight-binding model \cite{cite7}. This approach is distinct from the second moment method. In the $sp-d$ basis set, the surface energies can be derived with ease by respecting a charge neutrality rule  applied exclusively to the $d$-electrons, which contributes more to the cohesive energy. The $sp$-band is not subject to the charge neutrality. This charge neutrality results in a linear muffin-tin orbital charge distribution. The stoner model in this formalism has been demonstrated to yield accurate surface energies \cite{cite5}. The present study proposes a correction to the SMA potential and provides a more accurate interatomic potential. Additionally, it examines the charge transfers and magnetic properties.

\section{Interatomic potentials}
\subsection{Second moment approximation}
The second moment approximation is a semi-empirical interatomic potential that is employed to describe transition metals and alloys. The SMA utilizes an analytic expression to estimate the total energy:

$$ E_\text{tot} (r) = E_{r} (r) +  E_{b} (r),$$
 
where $E_{b} (r)$ is the attractive part related to the binding energy or the sum of Kohn-Sham eigenvalues:
 
$$ E_b(r) = -\left\lbrace  \sum \zeta^2 \exp\left[  -2q \left( \frac{  r_{ij}  }{r_0} -1  \right)  \right] \right\rbrace ^{1/2}, $$  

and the repulsive part $E_{r} (r)$ given by: 
$$ E_r(r) = \sum  A \exp\left[ \left( \frac{  r_{ij}  }{r_0} -1  \right)  \right] $$

$q$, $\zeta$, $A$ are parameters employed to fit the SMA potential to the potential of a higher level of theory. $r_0$ is the  equilibrium distance. This approximation is often limited to a single band, which is intended to describe the $d$-band. The surface energies obtained from this method are half of the surface energies obtained from a DFT calculation.

\subsection{Correction of the localized $d$-band approximation}
In a separate study, an analogous interatomic potential was derived to examine the impact of $sp-d$ hybridization  \cite{cite8}. The surface energies obtained are consistent with the results of an \textit{ab initio} calculation.

\begin{figure}[!h]    
 	\centering
		\includegraphics[width=0.45\textwidth]{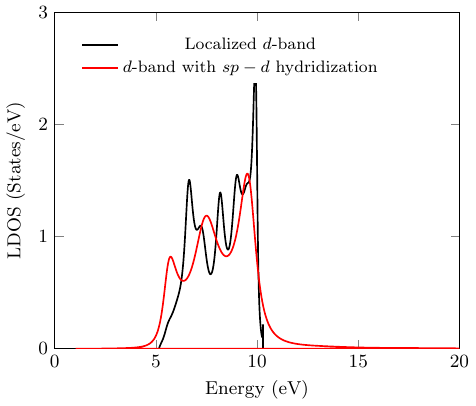}	
		\caption{Comparison of the localized $d$-band and $d$-band with the $sp-d$ hydridization.}      
		\label{fig:doslocdel}
\end{figure}		
 		
\begin{figure}[!h]
    \includegraphics[width=0.45\textwidth]{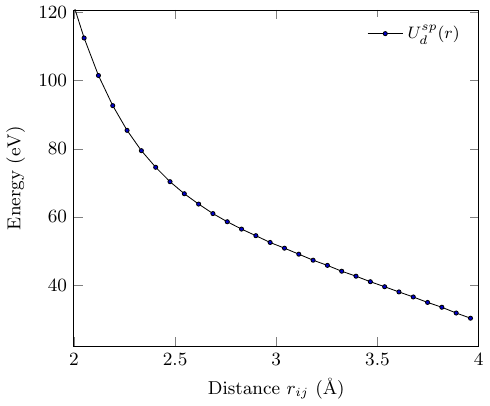}	
	\caption{ Variation of $U_d^{sp}$ with the distance. }      
	\label{fig:dist_Usp}
\end{figure}

As demonstrated in  Fig. \ref{fig:doslocdel}, the $d$ local density of states (LDOS) in the absence of $sp-d$ hybridization results in a narrow bandwidth.  The $d$ electrons are subsequently more localized compared to the $d$-band with the $sp-d$  hybridization (described as a delocalized $d$-band). $U_d^{sp}$ is employed to denote the correlation energy between the localized $d$-band and the delocalized $d$-band used as a correction \cite{cite4}.  This correction can be quantized approximately as the difference between the band energies of the localized and delocalized $d$-LDOS. 

$$ U_d^{sp} \approx \int_0^{E_f} n_d(E)^\text{loc.}E dE - \int_0^{E_f} n_d(E)^\text{deloc.}E dE, $$
  
where $n_d(E)$ is the $d$-LDOS. The $U_d^{sp}$ correction is always positive. Consequently, the delocalized $d$-band is always lower in energy than the localized $d$-band. The delocalized $d$-band is thus identified as the ground state of a transitional metal. The electronic structure of the $sp$-band is dependent on the interaction between one atom and its neighbors. Therefore, the $U_d^{sp}$ correction is dependent on the interatomic distance (Fig. \ref{fig:dist_Usp}).\\

\begin{figure}[!h]
		\includegraphics[width=0.45\textwidth]{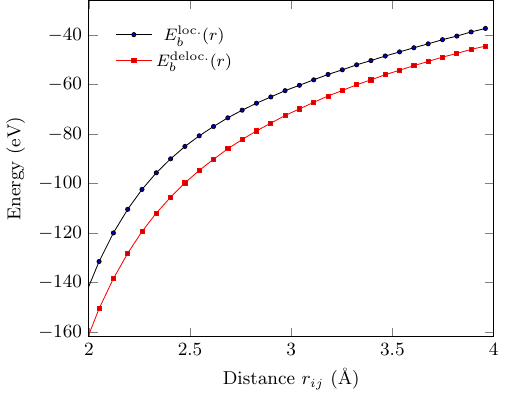}	
		\caption{Comparison of the attractive band energy in the localized $E_{b}^\text{loc.}$ and delocalized $E_{b}^\text{deloc.}$  approximations.}      
		\label{fig:pot_att_loc_del}
\end{figure}

The $d$-band with the $sp-d$ hybridization is consistently lower in energy compared to the localized $d$-band (Fig. \ref{fig:pot_att_loc_del}). The $U_d^{sp}$ value decreases as the distance between atoms gets smaller. This is because the overlap of the $sp$ and $d$ bands decreases as well. The use of a localized band is a valid approximation within the limit when the interatomic distance is large $r>>$. The localized d-band, excluding the $sp-d$ hybridization, is inadequate for accurately representing the electronic structure of transitional metals in proximity to equilibrium.  The following correction is required for the localized $d$-band potential: 

$$ E_{tot} (r) = E_{r} (r) +  E_{b} (r) - U_d^{sp}(r)$$

However, the relationship between $U_d^{sp}$ and the interatomic distance is not a trivial function. Consequently, this correction exhibits a greater degree of complexity than the use of the $sp-d$ hybridization.

\subsection{Case of the noble metals}
Regardless of whether the SMA incorporates the $sp-d$ hybridization or the previous correction, the description of the noble metal will be inaccurate. This is due to the fact that the band energy of noble metals, such as platinum, does not have an exponential shape as a function of interatomic distance.

\begin{figure}[!h]    
	\centering
	\begin{subfigure}[b]{0.25\textwidth}
        \includegraphics{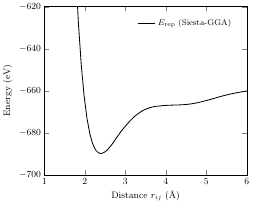}		
		\caption{Repulsive energy }      
		\label{fig:ebandpt}
	\end{subfigure}%
	~ 
	\begin{subfigure}[b]{0.25\textwidth}
		\centering
        \includegraphics{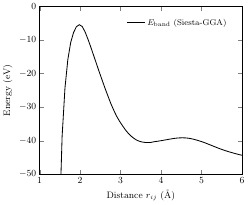}	
		\caption{Band energy }
	\end{subfigure}%
	\caption{Band energy (E$_\text{band}$) and repulsive energy (E$_\text{rep}$)  of  fcc Pt (Siesta-GGA).}
	\label{fig:ebandpt1}
\end{figure}

\begin{figure}[!h]
	\centering
	\includegraphics[width=0.45\textwidth]{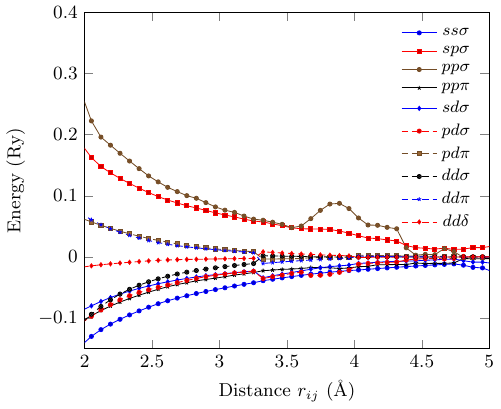}       
	\caption{Variation of hopping parameters with distance through the fitting of band structures.}
	\label{fig:evol_param1}
\end{figure}

As illustrated in Fig. (\ref{fig:ebandpt1}), the band energy and the repulsive energy (E$_\text{tot}$ - E$_\text{band}$) exhibit a bell-shaped profile. This shape cannot be represented by an exponential function, even if a fit is performed over a small interval near the equilibrium ($r_0$ = 2.77 \AA ${ }$ for platinum).  Therefore, it is not possible to have an \textit{ab initio} potential fit using the second or n-th moment restricted in the d-band because all the hopping parameters depend on distance (Figure \ref{fig:evol_param1}).

\subsection{An accurate $sp-d$ tight-binding potential}
In order to reproduce the behavior of hoping parameters with interatomic distances (Figure \ref{fig:evol_param1}), a linear combination of the exponential functions should be used:

$$
	(s,p,d)(s,p,d)(\sigma \pi \delta) =  \sum_\lambda   a_\lambda  \exp(b_\lambda \frac{r_{ij}}{r_0} )  
$$ 

Where $a_\lambda$ and $b_\lambda$  are the parameters.  $(s,p,d)(s,p,d)(\sigma \pi \delta)$  represents the hopping parameters: $ss\sigma, sp\sigma, sd\sigma, pp\sigma$ $pp\pi,pd\sigma, pd\pi  , dd\sigma, dd\pi, dd\delta $.  For the sake of simplicity, the model is constrained to two exponentials:

$$
(s,p,d)(s,p,d)(\sigma \pi \delta) =    a_1  \exp(b_1 \frac{r_{ij}}{r_0} ) +    a_2 \exp(b_2 \frac{r_{ij}}{r_0} )   
$$

\begin{figure}[!h]
	\centering
	\includegraphics[width=0.45\textwidth]{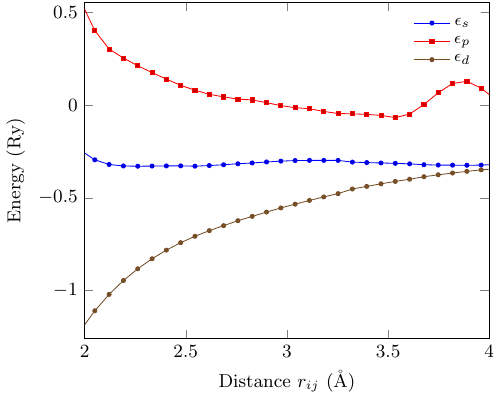}         
	\caption{Variation of the atomic energies with the distance through the fitting of band structures.}
	\label{fig:evol_param2}
\end{figure}

The parameters $a_1$, $a_2$, $b_1$, and $b_2$ are determined through a process of  fitting, whereby the hopping parameters are adjusted with the curves depicted in Fig. (\ref{fig:evol_param1}). Despite the fact that the hopping parameters are dependent on the interatomic distance, an accurate description of the atomic energies is only possible with a dependency on interatomic distances, as demonstrated in Fig. (\ref{fig:evol_param2}). 

The second moment approximation does not account for the variation in atomic energies with distance, resulting in an inaccurate depiction of atomic behavior. The atomic energies $\epsilon_s$, $\epsilon_p$ and $\epsilon_d$ are hereby defined as follows:

$$
\epsilon_{s, p, d} = a_1  \exp(a_2 (r/r_0)) + a_3 \exp(a_4 (r/r_0))  
$$

\begin{figure}[!h]
	\centering
	\includegraphics[width=0.45\textwidth]{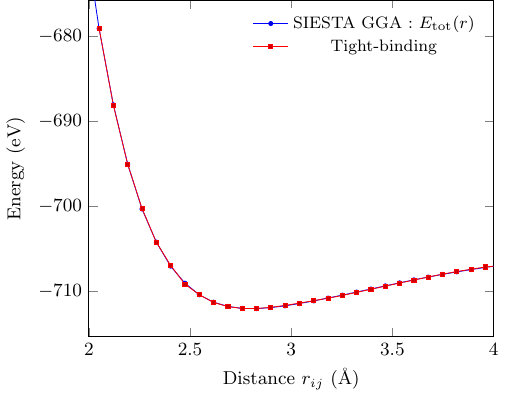}  	
	\caption{Interatomic potential for fcc platinum  obtained using hopping parameters and atomic energies as a function of interatomic distance.}
	\label{fig:Pot_pt}
\end{figure}

The variation of $\epsilon_s$ with the interatomic distance is nearly negligible. The potential obtained in this study is comparable to the potential obtained through DFT calculations (Fig. \ref{fig:Pot_pt}). The semi-empirical interatomic potential is derived by incorporating a repulsive contribution in the form of a polynomial function. There is an accurate long-range agreement between the semi-empirical and the DFT interatomic potential. This potential encompasses the correct electronic structure. 

\section{Charge distribution in transition metal surfaces and alloys}

According to DFT calculations in a linear combination of atomic orbitals (LCAO) basis set, the charge is neutral in transition metals and alloys per atom, per orbital, and per chemical species in the bulk and at the surface \cite{cite9}.  The charge neutrality derived from the Mulliken population is not consistent with a Bader charge analysis and experimental results \cite{cite10}. \\

\subsection{Charge transfer in alloys}
In the alloy L1$_0$ CoPt using a plane-wave basis set calculation, a charge transfer of $0.25$ electron from cobalt to platinum is observed, as determined by Bader charge analysis. A subsequent calculation of the other phases of CoPt reveals a decrease in charge transfer with increasing concentrations of Co. When the tight-binding approximation is restricted to the localized $d$-band, a charge neutrality is required. However, it is important to note that the charge neutrality observed in this specific context cannot be extrapolated to a global scale.  The quantification of charge transfer in the tight-binding approximation is achieved by filling the LDOS to obtain the sum of the valence electrons of the two species, Co and Pt, at the Fermi level:

$$ N_e = \int_{-\infty}^{E_f} \left[  n_{spd}^\text{Co}(E) + n_{spd}^\text{Pt}(E)\right]  dE   $$

%
The hopping parameters derived from an all-electron DFT calculation \cite{cite11} are employed to compute the partial charge in the $d$-band and to estimate charge transfers from Co to Pt, which is approximately 0.20 electrons. This estimation is in accordance with the Bader charge analysis and the experimental results \cite{cite10}. It can be posited that charge transfer can be calculated within the $sp-d$ tight-binding approximation.

\subsection{Charge transfer at the surface}
It was theorized that surface charge neutrality would align with bulk charge neutrality. The implementation of charge neutrality per orbital results in a discrepancy between the obtained surface energies and those calculated by DFT. Despite the existence of a charge neutrality rule, the electron distribution should be considered. Therefore, the hypothesis is proposed that charge neutrality is only applicable at the surface for electrons in the $d$-band. The charge neutrality rule is implemented as follows: due to the constraint imposed on the atoms at the surface to maintain charge neutrality, the atomic levels of the $d$-electrons are  shifted \cite{cite13}. \\

A shift on the $d$ atomic level $\delta \epsilon_D$ is applied. The selfconsitency treatment will be $ \epsilon_D = \epsilon_D + \delta \epsilon_D $ until the charge in the $d$-band is conserved at the surface \cite{cite11_0}. It is evident that the $s$ and $p$ atomic energies  ($\delta \epsilon_s = 0 $ and  $\delta \epsilon_p =0  $) are  unshifted.  Subsequent to this procedure, delocalized $sp$ electronic charge arises at the Fermi level $(S+1)$ in consistent with the charge distribution obtained in a linear muffin-tin orbital calculation \cite{cite5}.

\section{Stoner magnetism}
\begin{figure}[!h]
	\centering
	\includegraphics[width=0.45\textwidth]{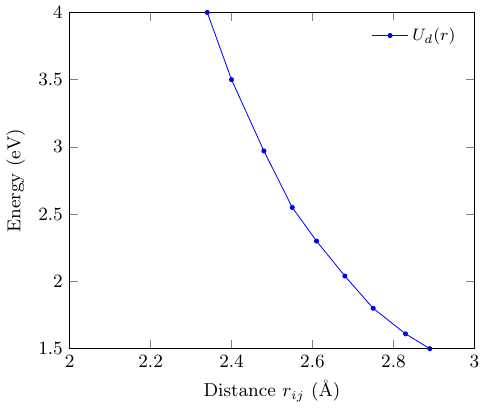}  
	\caption{Variation of  $U_d$  as a function of interatomic distance for a $d$-band without the $sp-d$ hybridization. }
	\label{fig:udist}
\end{figure}

The local Hubbard hamiltonian, when constrained within the $d$-band in the mean field approximation, yields the following expression for the spin magnetic moment \cite{cite4}: 

$$ \mu = \frac{5}{U_d}\Delta \epsilon  = \frac{\Delta \epsilon }{I}, $$ 

where $U$ is the Hubbard parameter, $I$ is the Stoner parameter,  $\Delta \epsilon$ is the exchange splitting. The variation of energy due to magnetization, which is negative for ferromagnetic elements, can be expressed as:
$$
	\Delta E_\text{mag} = \Delta E_{b} -  \frac{1}{20}U_d\mu^2
$$

\[ 
\begin{tabular}{ccc}
$n_\uparrow(E) =  n(E-\frac{\Delta \epsilon }{2})$ \\
$n_\downarrow(E) =  n(E+\frac{\Delta \epsilon }{2})$ \\
\end{tabular}
\]
The spin magnetic moment is given by: 

$$
	\mu = \int_{-\infty}^{E_f} n(E-\frac{\Delta \epsilon }{2}) dE - \int_{-\infty}^{E_f} n(E+\frac{\Delta \epsilon }{2}) dE
$$
This expression is adjusted with the expression of the spin magnetic moment using the Stoner model $\mu  = 5 \cdot \Delta E / U_d $ in order to obtain the DFT or experimental magnetic moment and  $U_d$  \cite{cite11_0}. In the context of the fcc Cobalt in a  localized $d$-band approximation, $\Delta E_{mag} >0 $ and  $U_d=3$ eV \cite{cite4}. $U_d$ is found to be significantly influenced by the interatomic distance (Fig. \ref{fig:udist}). 

The application of the correction $U_d^{sp}$ is imperative to ensure the attainment of an accurate description of the binding energy. When the $sp-d$ hybridization is included, the resultant value of $U_d$ is 6 eV. This value encompasses all the necessary corrections. The magnetic interatomic potential can be obtained using this single parameter, $U_d$. In the event that the $sp-d$ hybridization is disregarded, it is imperative to incorporate the correction $U_d^{sp}$ into the magnetic energy calculation.

\section{Conclusion}
The $d$-band approximation, excluding the $sp-d$ hybridization, necessitates multiple corrections to accurately describe the physical properties of transitional metals and alloys. The correlation between this approach and the $d$-band with the $sp-d$ hybridization is established through the utilization of a single parameter that lacks an analytic expression. The parameter in question has been demonstrated to be effective in correcting the SMA interatomic potential in instances where the exponential function can be employed as an approximation. It has been demonstrated that a more accurate semi-empirical potential can be obtained by utilizing hopping parameters depending on the intatomic  distances. The $d$-band approximation, excluding the $sp-d$ hybridization, is frequently used with a charge neutrality rule per chemical species in alloys. This approach stands in contrast to the findings derived from first-principles calculations, which indicate an occurrence of charge transfer. It is evident that charge transfers cannot be calculated by this level of theory. The subsequent correction increases the complexity of the method. It is imperative to recalibrate the magnetic ground state within the $d$-band approximation, excluding the $sp-d$ hybridization. The implementation of these corrections in all cases is not feasible and universally transferable across all physical environments.

\bibliography{index}

\end{document}